\documentclass[pre,preprint]{revtex4}
\usepackage{graphicx}
\usepackage{amsmath}

\def\eq#1{eq.~(\ref{eq:#1})}
\def\fig#1{fig.~\ref{fig:#1}}
\def\Fig#1{Fig.~\ref{fig:#1}}

\def\kt{k_BT}
\def\s{\,\mathrm{s}}
\def\fs{\,\mathrm{frames/s}}
\def\Hz{\,\mathrm{Hz}}
\def\nm{\,\mathrm{nm}}
\def\pN{\,\mathrm{pN}}
\def\pNs{\,\mathrm{pN/s}}
\def\mum{\,\mathrm{\mu m}}
\def\pNnm{\,\mathrm{pN/nm}}
\newcommand{\avg}[1]{\langle #1\rangle}

\begin{document}

\title{Dynamic force spectroscopy on multiple bonds:\\
experiments and model}
\author{T. Erdmann}
\affiliation{FOM Institute for Atomic and Molecular Physics, Kruislaan 407, 1098 SJ Amsterdam, The Netherlands}
\author{S. Pierrat}
\affiliation{Laboratoire de Physico-Chimie Curie, Institut Curie, F-75005 Paris, France}
\author{P. Nassoy}
\affiliation{Laboratoire de Physico-Chimie Curie, Institut Curie, F-75005 Paris, France}
\author{U.~S. Schwarz}
\affiliation{University of Heidelberg, Bioquant 0013, Im Neuenheimer Feld 267, D-69120 Heidelberg, Germany}

\begin{abstract}
  We probe the dynamic strength of multiple biotin-streptavidin
  adhesion bonds under linear loading using the biomembrane force
  probe setup for dynamic force spectroscopy.  Measured rupture force
  histograms are compared to results from a master equation model for
  the stochastic dynamics of bond rupture under load. This allows us
  to extract the distribution of the number of initially closed bonds.
  We also extract the molecular parameters of the adhesion bonds, in
  good agreement with earlier results from single bond experiments.
  Our analysis shows that the peaks in the measured histograms are not
  simple multiples of the single bond values, but follow from a
  superposition procedure which generates different peak positions.
\end{abstract}

\maketitle

\section{Introduction}

Cell adhesion in multicellular organisms is mediated by highly
specific, but weak receptor-ligand bonds
\cite{a:Bell1978,a:Bongrand1999,c:evan07}. Typical bond energies are
in the range of several $\kt$ so that bond lifetimes are finite due to
thermal activation from the environment. Since cellular adhesion sites
usually operate under mechanical load, different experimental
techniques have been developed to study rupture of single biomolecular
bonds under an applied load. Motivated by \emph{atomic force
  microscope} (AFM) measurements on biotin-streptavidin
\cite{a:FlorinMoyGaub1994}, it has been shown that the force at which
a bond breaks under an applied load is a stochastic variable with a
probability distribution which depends on the loading protocol and
which can be explained by Kramers theory for the escape of an
overdamped particle from a potential well over a sharp transition
state barrier \cite{a:EvansRitchie1997,c:seif98}. This insight led to
the new field of \emph{dynamic force spectroscopy} (DFS)
\cite{a:Merkel2001}.  For the biotin-streptavidin system,
the concept of dynamic bond strength has been confirmed by experiments
with the \emph{biomembrane force probe} (BFP) \cite{a:MerkelEtAl1999}.
If force $F$ increases linearly in time $t$ with loading rate $m$,
i.e.~if $F = mt$, a rupture force distribution results in which the
most frequent rupture force $F^*$ is proportional to the logarithm of
loading rate (the same result has been found for the average rupture
force \cite{c:tees01}). Dynamic force spectra use this relation to
chart the binding landscape: a straight line in the plot of $F^*$
versus $\ln m$ is characteristic for a single dominant barrier
limiting escape. The slope is determined by the distance of the
barrier from the ground state (the \emph{reactive compliance} $x_b$)
and extrapolation to $F^* = 0$ yields the \emph{unstressed
  off-rate} $k_0$ of the bond over this barrier. A series of energy
barriers reveals itself as a sequence of straight lines with
increasing slope, thus defining different regimes, each dominated by
one transition state barrier.

Single bond DFS experiments require a very low frequency of successful
binding events to avoid multiple attachments. This renders single bond
experiments time consuming and accumulation of rupture force
histograms difficult. In the following, we describe DFS-experiments
with the BFP on the biotin-streptavidin system at four different
loading rates, in which we allow formation of multiple bonds.
Experimentally, the streptavidin-biotin system is the best studied
example for molecular bonding. In detail, it has been investigated
with AFM \cite{a:FlorinMoyGaub1994}, BFP \cite{a:MerkelEtAl1999} and
flow chambers \cite{c:pier02}. In the latter case, multiple bonds have
been probed, but a persistent problem with flow chambers is that force
is distributed in a heterogeneous way over the different bonds, thus
rendering a quantitative analysis difficult.  In contrast, force
distribution is expected to be significantly more homogeneous for the
BFP as used here. The higher frequency of binding events introduced
here makes acquisition of rupture force histograms considerably
easier. However, in order to evaluate the histograms, a theoretical
model is required which allows to extract the single bond properties
from the multiple bond data.

Deterministic models for the rupture of multiple, parallel bonds have
been analysed before for constant \cite{a:Bell1978} as well as linear
loading \cite{a:Seifert2000}. While deterministic models describe the
average number of bonds and are most appropriate for large systems,
stochastic models are required for small systems with a finite number
of bonds. Here the simplest possible case is the irreversible rupture
of a small number of equivalent bonds, which can be described in the
mathematical framework of a one-step master equation \cite{c:tees01}.
Multiple bonds also allow rebinding of broken bonds, which can be
described in the same framework
\cite{a:ErdmannSchwarz2004a,a:ErdmannSchwarz2004c}.  It has also been
applied before to the case of linear loading commonly used in DFS
\cite{a:ErdmannSchwarz2004b}.  Similar approaches have been also used
to describe DFS on titin, where the different bonds may
correspond to different hydrogen bonds within a single Ig27-domain
\cite{maka01} or to the different Ig27-domains within titin
\cite{c:brau05}.

In this paper, we use the framework of a one-step master equation to
evaluate data from multiple bond DFS on the well-established
streptavidin-biotin system. Because multiple bonds are now allowed in
the experimental setup, the exact number of initial bonds in each
experiment is a stochastic variable. By comparing numerical solutions
of the master equation to experimental histograms, we can estimate the
corresponding probability distribution. In this way, we show for the
first time how the dissociation spectrum of multiple bond DFS results
from the superposition of the contributions from different numbers of
initially closed bonds.

\section{Experiments}

\begin{figure}
\includegraphics[width=0.5\textwidth]{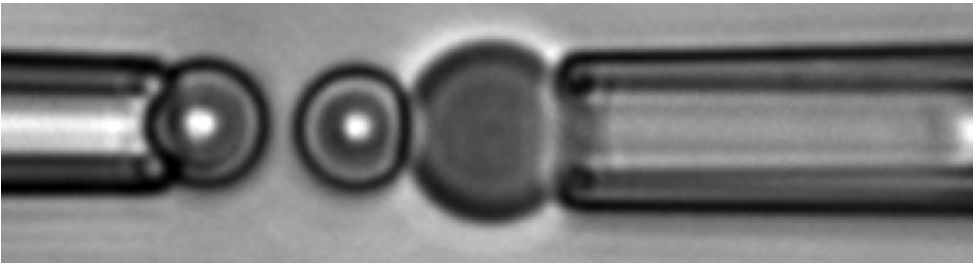}
\caption{Videomicrograph of the biomembrane force probe (BFP). The red
blood cell (on the right) acts as an elastic element, which
transforms 
displacement of the pipette into force exerted on the binding site. The
probe bead (attached to the red blood cell) and the test bead (on the
left) are glass spheres ($3 \mum$ in diameter) decorated with
complementary receptor and ligand molecules, in this case streptavidin
and biotin.}
\label{fig:setup}
\end{figure}

To monitor rupture forces between multiple streptavidin-biotin bonds at
different loading rates, we used the BFP-instrument as described
previously \cite{a:MerkelEtAl1999} and depicted in \fig{setup}. In
brief, for use as force transducers, biotinylated red blood cells were
pressurized into a spherical shape with a micropipette. The suction
pressure sets the red cell membrane tension, and thus the BFP spring
constant, $k_F$, in a tunable fashion at values as low as $0.1-1 \pNnm$.
Probe and test glass microspheres were decorated respectively with
streptavidin and biotin following a common procedure, in which
polyethylene glycol (PEG) crosslinkers (Nektar) were used to inhibit
non-specific adhesion\cite{a:PerretEtAl2002}. The streptavidin-coated
probe bead was firmly attached to the apex of the biotinylated red blood
cell capsule. The biotin-coated test bead was maintained in a second
micropipette and manoeuvred to/from the BFP probe at controlled
impingement forces ($ < 25 \pN$) and retraction speeds by precision
piezo-translators. Bead positions were determined by on-line video
processing at a sampling rate up to $180\ \fs$ and with a spatial
resolution of about $6\ \nm$. Validation of our set-up was achieved by
measuring rupture forces of individual streptavidin-biotin bonds for
loading rates between $5$ and $50,000 \pNs$ (data not shown, see
\cite{t:Pierrat2004} for details). The obtained dynamic force spectrum,
which collects the most probable unbinding force for each loading rate,
was in excellent agreement with previous reports
\cite{a:MerkelEtAl1999}. In particular, the derived energy landscape
exhibits two transition barriers characterised by two reactive
compliances, $x_{b,1} = 0.14 \nm$ and $x_{b,2} = 0.51 \nm$.

\begin{figure}
\includegraphics[width=0.5\textwidth]{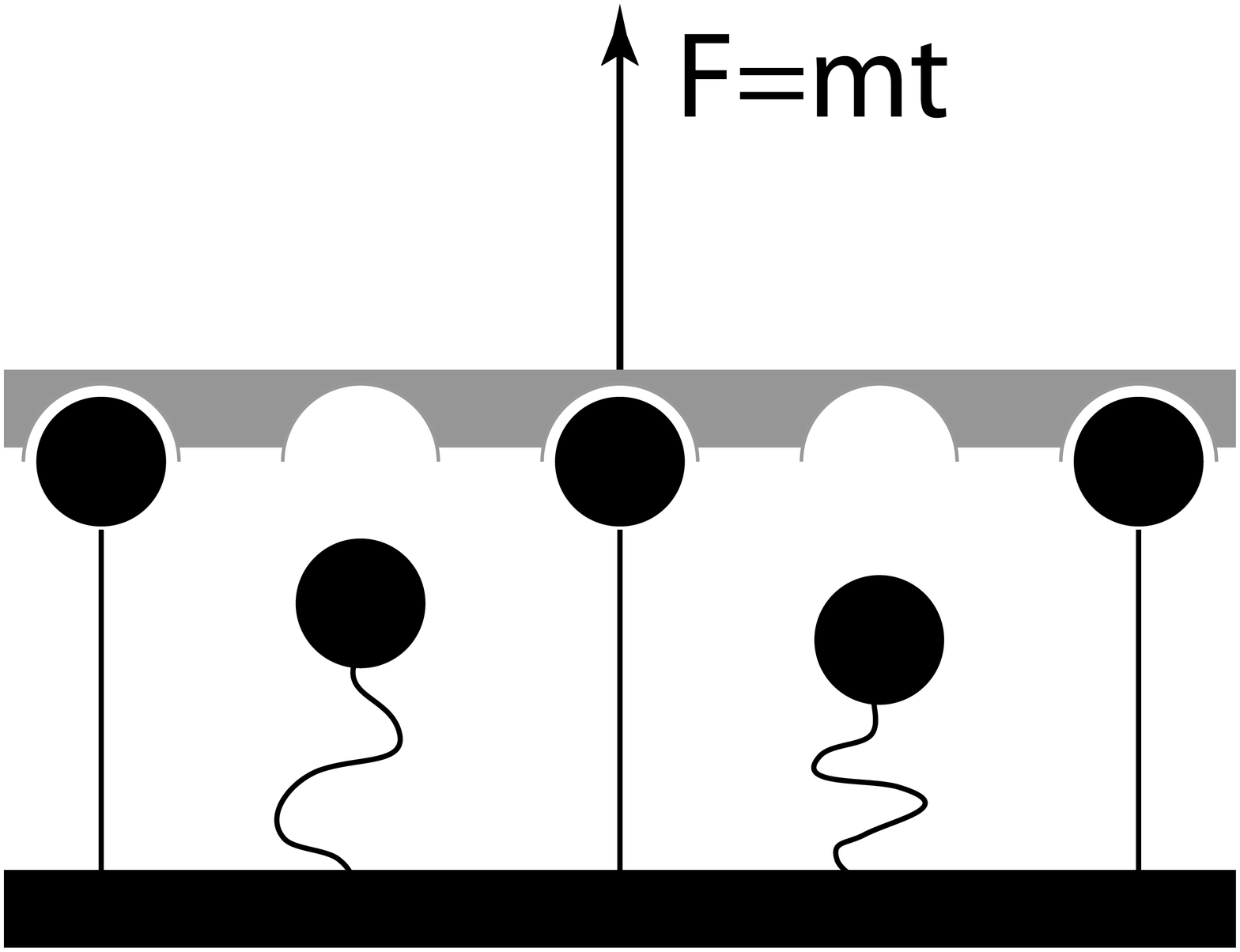}
\caption{Schematic representation of an adhesion cluster in the DFS
experiment from \fig{setup}. Between substrate and transducer $N_t$
receptor-ligand bonds were established (here $N_t=5$) of which $i$ are
still closed (here $i=3$) while $N_t-i$ have already ruptured (here
$N_t-i=2$). The load $F$ applied to the transducer is assumed to be
equally distributed over the $i$ closed bonds, thus each of them feels
the force $F_b = F/i$. In our experiments, force is increased linearly
in time with loading rate $m$, i.e.~$F = mt$.}
\label{fig:cartoon}
\end{figure}

As compared with single bond DFS, we mainly implemented two changes in
the experimental procedure. First, while an attachment frequency of $1$
per $10$ touches is the usual criterion that ensures high probability of
single bond formation, we increased the surface density of specific
sites on the glass beads to reach an adhesion frequency of about $1$ per
$3$ touches. According to Poisson statistics, this increases the
likelihood of picking multiple bonds. Second, and more important, since
our goal was to quantitatively account for the distribution of all
measured rupture forces, good accuracy for both low and high forces was
required. Despite its advantages of tunability and softness, the red
blood cell-based transducer is only linear for extensions below $\sim
0.5 \mum$ \cite{a:SimsonEtAl1998}. In practice, this led us to set the
BFP spring constant around $1 \pNnm$ and the range of explored loading
rates between $100$ and $5000 \pNs$. In consequence, we focused on the
regime dominated by the outer barrier of the streptavidin-biotin energy
landscape (with a reactive compliance $x_{b,2} = 0.51 \nm$).

For biotin-streptavidin, rupture experiments with different
experimental techniques indicated a dependence of unbinding forces on
binding history, namely on the contact time that is available for
binding before loading begins. In \cite{a:PincetHusson2005} this was
assigned to the existence of an additional energy minimum (the
absolute ground state) which is reached only after a large contact
time and is more stable against rupture than the intermediate ground
state that is reached first. In our experiments, we use the same
contact protocol as in \cite{a:MerkelEtAl1999} with short contact
times of $2-3 \s$, thus we expected the bonds to unbind from the
same initial state probed before.

\section{Model}

\begin{figure}
\includegraphics[width=0.8\textwidth]{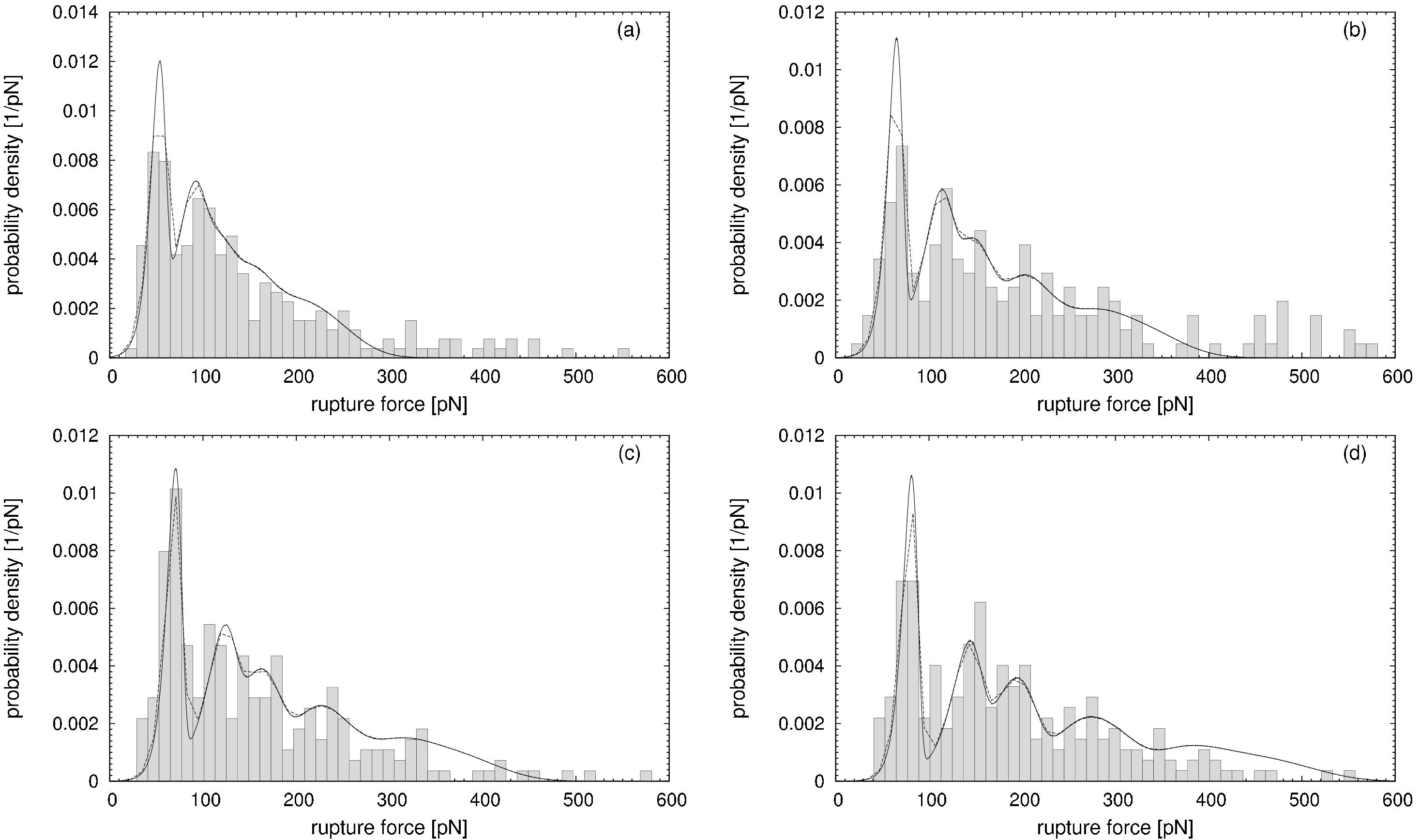}
\caption{Rupture force histograms for loading rates (a) $m = 130$, (b)
  $615$, (c) $1250$ and (d) $5000 \pNs$ which include $220$, $170$,
  $230$ and $228$ attachment events, respectively. The shaded bars
  show experimental results which were collected in bins of
  approximately $12 \pN$ width. The histograms are normalised in
  respect to the total number of events and the bin width, so that
  they show the probability density of events. The solid lines are the
  simulation results fitted simultaneously to the complete data set
  (a-d). For the fitting procedure, the numerical results were
  discretized. These discretized curves are shown as dashed lines. The
  optimal fit parameters are $k_0 = 0.02 \Hz$ and $F_0 = 7.83 \pN$.}
\label{fig:histogram}
\end{figure}

\Fig{cartoon} shows a schematic representation of the experimental setup
from \fig{setup}. Between two opposing surfaces $N_t$
receptor-ligand pairs are arranged in parallel. Each receptor can either
be bound to its ligand (closed bond) or unbound (open bond). At a given
time $t$ there are $i$ closed and $N_t-i$ open bonds. For the large
loading rates that we use in our experiments, it is sufficient to
consider irreversible bonds which cannot rebind after rupture
\cite{a:Seifert2000,a:ErdmannSchwarz2004b}. Thus, the number $N_t$ of
bonds which is relevant for the rupture process is the number of closed
bonds that has formed prior to loading. Rupture of a bond occurs
stochastically through thermal activation. Following Bell
\cite{a:Bell1978} the off-rate for rupture of a bond increases
exponentially with the force $F_b$ exerted to it, i.e.~$k_{off} = k_0
\exp(F_b / F_0)$. Here $k_0$ is the unstressed off-rate at vanishing
force and $F_0 = \kt / x_b$ is the intrinsic force scale for bond
rupture, which is set by thermal energy $\kt = 4.1 \pN\nm$ and reactive
compliance $x_b$.

An essential part of the modelling is a reasonable assumption on how
force is distributed over the closed bonds in a cluster. Here we assume
that the force $F = mt$, which is determined by the displacement of the
red blood cell, is independent of the number of closed bonds and is
shared equally between them, i.e.~$F_b = F/i$. In our setup, this
assumption is appropriate because the bonds are attached to a solid
support of finite curvature.  Because the binding times of tethered
polymer bonds increase very rapidly with the separation of the opposing
surfaces \cite{a:JeppesenEtAl2001}, our short contact times make
formation of bonds at large separations unlikely. Moreover biomolecular
bonds carried by extended polymer tethers dissociate rapidly under force
\cite{uss:erdm07a}. Therefore we expect the relevant bonds to form
mainly within a narrow range of surface separations corresponding to the
area of closest approach.  Because the tethers that attach the ligands
to the solid support have a contour length ($\sim 23.2 \nm$) which is
small compared to the extension of the soft transducer, unbinding of a
tether has a negligible effects on transducer extension and force.

The number of closed bonds $i$ in a cluster reduces with time from $i =
N_t$ (initial, bound state) to $i = 0$ (final, dissociated state).
Because for shared loading all bonds are equivalent, the probabilities
$p_i(t)$ ($i=0 \dots N_t$) to find $i$ closed bonds at time $t$
completely characterise the system. Their time evolution follows the
one-step master equation
\begin{equation}\label{eq:MasterEquation}
\frac{dp_i}{dt} = r_{i+1} p_{i+1} - r_i p_i\,.
\end{equation}
This equation states that $i$ decreases through rupture of a closed bond
with the rate $r_i = i k_{off} = i k_0 \exp(mt / i F_0)$, where the
linear factor $i$ accounts for the fact that at any given time, each of
the $i$ remaining bonds can be the next to break. Once the master
equation is solved, one can calculate the cluster dissociation rate
$D_{N_t}(t) = \dot p_0(t) = r_1 p_1(t)$ as a function of time, which can
be converted to $D_{N_t}(F)$ using the linear relation $F = mt$ between
time and force.

\begin{figure}
\includegraphics[width=0.8\textwidth]{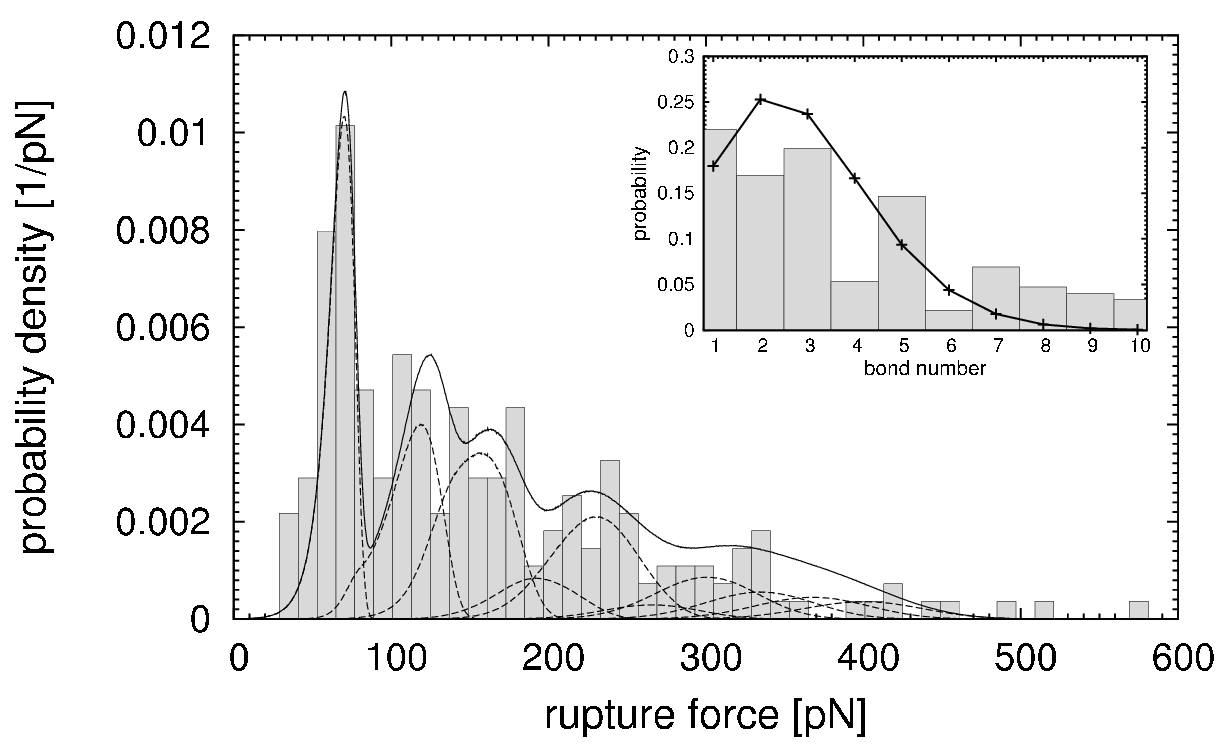}
\caption{Superposition of the different $D_{N_t}(F)$ (dashed lines)
  resulting in the final dissociation probability $D(F)$ (solid line)
  fitting the experimental histogram for loading rate $m = 1250 \pNs$.
  The different $D_{N_t}(F)$ are weighted with the distribution
  $p_{\alpha}(N_t)$ shown as a histogram in the inset.  The solid line
  with crosses is a least-mean-square fit to a Poisson distribution.}
\label{fig:decomposition}
\end{figure}

Our model contains three unknown parameters, which can be determined
through comparison with experiment: the unstressed off-rate $k_0$, the
intrinsic force scale $F_0$ (or, equivalently, the reactive compliance
$x_b$) and the initial number of closed bonds $N_t$. In our experiments,
the latter is itself a stochastic variable and cannot be controlled for
every rupture event. Instead we assume that $N_t$ follows a
probability distribution
\begin{equation}\label{eq:free}
p_{\alpha}(N_t) = \alpha_{N_t} / (\alpha_1 + \dots + \alpha_{N_t^m})\,.
\end{equation}
which is characterised by the set of $N_t^m$ coefficients $\alpha =
\{\alpha_{1}, \dots, \alpha_{N_t^m}\}$.  $N_t^m$ is the maximal number
of initial bonds considered and will be chosen in such a way that it
does not affect the outcome of our parameter estimation.  Because
experimentally we only record data involving deformation of the
transducer, events with $N_t = 0$ are not considered in the
theoretical analysis. If closed bonds formed at a constant rate over a
given time interval, $p_{\alpha}(N_t)$ would be a Poisson distribution
\begin{equation}\label{eq:poisson}
p_{\lambda}(N_t) =  \frac{\lambda^{N_t}e^{-\lambda}}{(1-e^{-\lambda})N_t!} \,,
\end{equation}
as has been reported before for similar experimental setups
\cite{c:pier02,a:Zhu2000}. The Poisson distribution contains the single
parameter $\lambda$ which determines the average number of broken bonds
as $\avg{N_t} = \lambda / (1-e^{-\lambda})$.

Experimental histograms are linear combinations of the distributions
$D_{N_t}(F)$ for single $N_t$ in which the distribution \eq{free}
determines the coefficients:
\begin{equation}\label{eq:histogram}
D(F) = \sum_{N_t=1}^{N_t^m} p_{\alpha}(N_t)D_{N_t}(F)\,.
\end{equation}
For shared loading with a time-dependent force, analytical solutions
for \eq{MasterEquation} do not exist \cite{a:ErdmannSchwarz2004b}.
Therefore we use the Gillespie algorithm \cite{mc:Gillespie1976} for
exact stochastic simulations to generate rupture trajectories with the
stochastic dynamics described by the master equation. For given values
of $k_0$ and $F_0$, distributions $D_{N_t}(F)$ of rupture force $F$
are calculated for $N_t = 1 \dots N_t^m$ at the four loading rates.
The distributions are superimposed via $p_{\alpha}(N_t)$
as in \eq{histogram} and the relative weights are
determined by a least-mean-square fit of $D(F)$ to the complete set of
experimental data. For this purpose, the simulation data are
discretized in the same way as the experimental data and every bin is
used as one data point. This procedure is iterated until values for
$k_0$ and $F_0$ are found that minimize the deviations between
simulations and experimental histograms.

\begin{figure}
\includegraphics[width=0.8\textwidth]{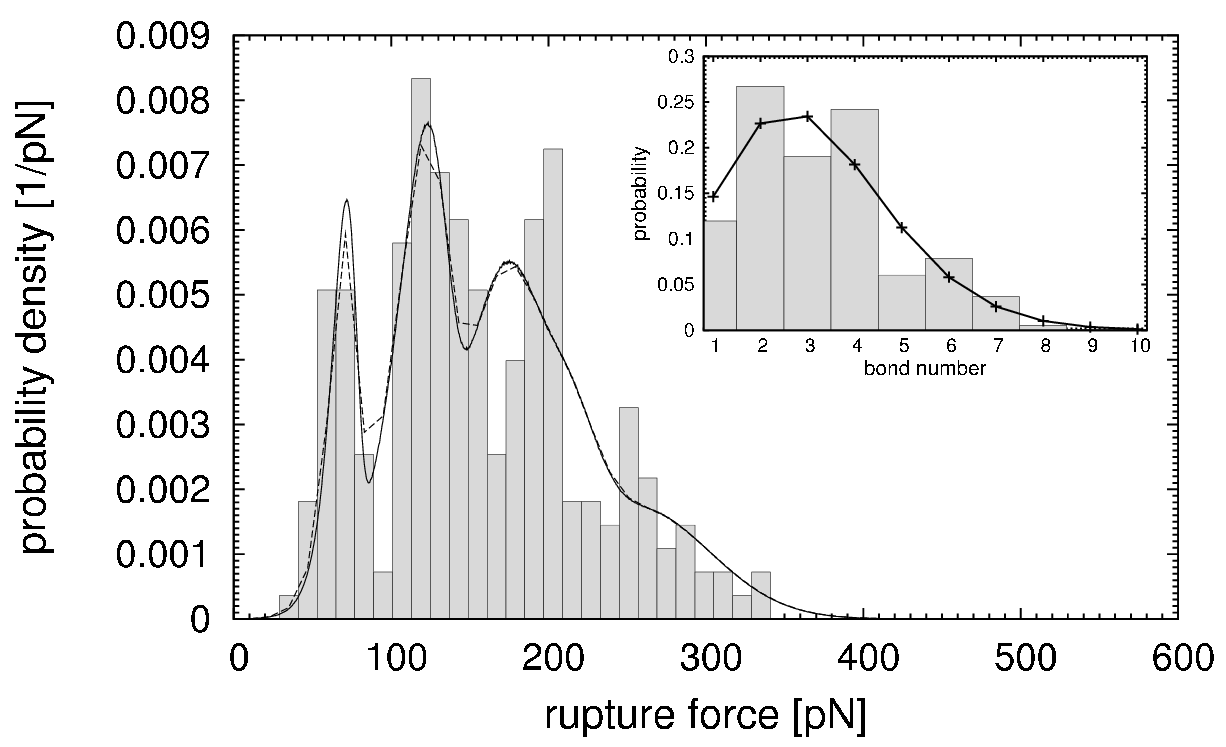}
\caption{Rupture force histogram with 230 data points and for loading
  rate $m = 1250$ pN/s simulated from the Poisson distribution from
  the inset of \fig{decomposition}. The lines and the inset
  represent the same analysis as done before for the experimental
  data.}
\label{fig:artificialdata}
\end{figure}

\section{Results}

\Fig{histogram} shows experimentally obtained rupture force histograms
including $220$, $170$, $230$ and $228$ rupture events for the four
different loading rates $m = 130$, $615$, $1250$ and $5000 \pNs$,
respectively. The histograms are compared to simulated rupture force
distributions which are calculated using $\sim 10^7$ rupture
trajectories at each value of $N_t$ and loading rate. Fitting
simulations to measurements yielded the unstressed off-rate $k_0 =
0.02 \Hz$ and the intrinsic force scale $F_0 = 7.83 \pN$,
corresponding to the reactive compliance $x_b = \kt/F_0 = 0.52 \nm$.
Thus our results extracted from multiple bond data compare very
favorable with earlier results from single bond experiments, which
gave $x_b = 0.51 \nm$ (a value for $k_0$ has not been given before).
Closer inspection of \fig{histogram} shows that the simulated
distributions approximate the positions and heights of the peaks in
the histograms in a reasonable way. Note that the fits were not done
separately to the different data sets for the different loading rates,
but simultaneously to the complete set of experimental data. Thus the
model is capable to fit well the whole range of loading rates probed.

\Fig{decomposition} shows for the case $m = 1250 \pNs$ how the
different rupture force distributions $D_{N_t}(F)$ combine to give
$D(F)$. Each curve $D_{N_t}(F)$ displays a single maximum followed by
a super-exponential decay at large forces. In general, their peaks
cannot be identified with the peaks in the experimental histograms as
it was done in previous analyses (e.g.~in
\cite{a:FlorinMoyGaub1994,a:ZhangMoy2003,c:aule04}). The best
agreement is found for $N_t = 1$, i.e.~between the peak of $D_1(F)$
and the first experimental peak. In contrast, the positions of the
peaks in $D_3(F)$ or $D_4(F)$ cannot be guessed from the histograms.
In general, the peak positions of the $D_{N_t}(F)$ do not increase
linearly with $N_t$.  Although this has been pointed out before
\cite{a:Seifert2000}, a \emph{force quantum} has occasionally been
discussed in the literature. Our analysis shows that the situation is
more complex and that accurate identification of the $D_{N_t}(F)$ from
the simulation data is required to extract the single bond data from
the multiple bond histograms.

\begin{figure}
\includegraphics[width=0.6\textwidth]{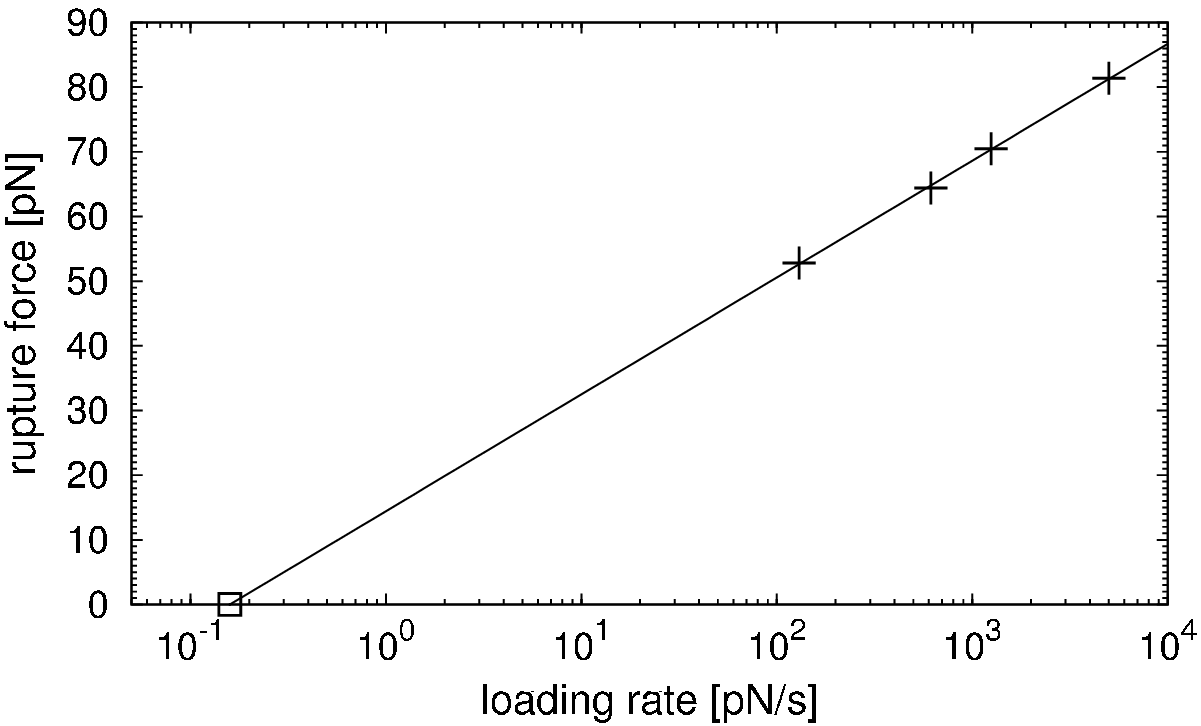}
\caption{Most frequent rupture force for single bonds as function of
loading rate. The crosses are the maxima of the fitted single bond
rupture force distributions $D_1(F)$ for the four different loading
rates. The values are $F^* = 52.8$, $64.4$, $70.5$ and $81.4 \pN$ for
increasing $m$. The \emph{dynamic force spectrum} reveals a straight
line with slope $F_0 = 7.85\pN$ which intersects $F^* = 0$ at $k_0 F_0 =
0.159 \pNs$ (marked by a square) so that $k_0 = 0.02 \Hz$.  The reactive
compliance is $x_b = \kt / F_0 = 0.52 \nm$.}
\label{fig:singlebond}
\end{figure}

The extracted probability distribution $p_{\alpha}(N_t)$ for the
number of initial closed bonds $N_t$ is the histogram shown as inset
in \fig{decomposition}.  Its average is $\avg{N_t} = 3.8$. The largest
$N_t$ included was $N_t^m = 10$, because larger $N_t^m$ did not make a
difference for the estimate of the molecular parameters.  The inset
also shows the result of a least-mean-square fit to a Poisson
distribution, which is characterized by $\avg{N_t} = 3.0$ ($\lambda =
2.8$). The reduced average reflects the fact that the Poisson
distribution underrepresents the tail of the distribution extracted
from the experimental data. However, it is also important to note that
the reconstruction at large forces is not very reliable due to the few
events in the experimental data (compare \fig{histogram}). In order to
test whether the Poisson distribution is compatible with our
experimental data, we generated artificial rupture histograms by
drawing random values for $N_t$ from the Poisson distribution exactly
as often as experimental data points were present.
\Fig{artificialdata} shows a corresponding histogram for loading rate
$m = 1250$ pN/s. The lines and the inset represent the same analysis
as done before for the experimental data.  The new Poisson
distribution is slightly different, with $\lambda = 3.1$ instead of
$\lambda = 2.8$, while the extracted values for $k_0$ and $F_0$ are
essentially unchanged. Taken together, the data simulation suggests
that the finite size of the data set can explain the deviations
between experimental data and model fit.

Figs.~\ref{fig:histogram} and \ref{fig:decomposition} suggest that the
most salient feature of both the experimental and simulation data is
the first peak, which corresponds to single bond rupture. In
\Fig{singlebond} we plot the fitted positions of the first peaks as a
function of $\ln m$.  As expected for dynamic force spectra, the data
points can be fitted to a linear curve with the slope $F_0 = 7.85 \pN$
and the intersection with $F^* = 0$ at $k_0F_0 = 0.159 \pNs$. This
again yields $k_0 = 0.02 \Hz$ and $x_b = 0.52 \nm$ for the unstressed
off-rate and the reactive compliance, respectively, demonstrating that
our analysis is consistent. These results suggest that the first peaks
in a series of multiple bond rupture force histograms at different
loading rates as shown in \fig{histogram} can be used to achieve a
quick and simple estimate of the molecular parameters. In particular,
this estimate can then be used as a starting point for the detailed
numerical analysis presented above.

\section{Discussion}

We have used multiple bond DFS-measurements with the BFP-setup to
probe the outer barrier of the biotin-streptavidin bond. If multiple
attachments are allowed, the frequency of successful binding events is
larger than in classical single bond experiments so that information
can be obtained more efficiently in the form of rupture force
histograms. As demonstrated here, analysis of these histograms
requires theoretical modelling of the rupture of parallel adhesion
bonds under an applied load because the peaks of the histograms follow
from a superposition procedure. Using an established stochastic model
for bond rupture under shared loading in combination with the
assumption of a variable number of initial bonds, we showed for the
first time how this superposition looks in practise, compare
\fig{decomposition}. The molecular bond parameters $k_0$ and $x_b$
were extracted in good agreement with previous results from single
bond experiments.

One particular feature of our approach is that with a simple model for
multiple bond rupture we arrived at an extraction of single bond data
which is both successful and efficient.  Our model for force
transduction and adhesion bonds is rather generic and can be easily
extended to include additional effects like finite transducer and
tether stiffnesses or generalised models for the binding landscape.
E.g.~it was shown for protein unfolding that analysis of histograms
can be used to identify the shape of the binding landscape and
identify deviations from the assumption of sharp transition barriers
underlying DFS \cite{a:SchlierfRief2006}. We expect that the agreement
between experimental and theoretical histograms can be improved by
including more substructure into the model. Recently, it has been
proposed that deviations in rupture force histograms from the DFS
predictions are caused by an intrinsic heterogeneity of bond
parameters \cite{a:RaibleEtAl2006}. The increased width and large
force tails could be described imposing Gaussian noise on $F_0$.
Although this effect will be small compared to the variations caused
by the distribution of $N_t$ and will not alter the basic results, it
might explain residual events at large rupture forces (another
possible explanation is that some initial population of the
biotin-streptavidin ground state exists after all).  Both examples
show that for detailed studies of adhesion bonds, a model-based
analysis of the histograms is required.

In this work, rebinding effects could be safely neglected. It has been
shown before that rebinding is only relevant if the dimensionless
loading rate $m/(k_0 F_0) < N_t$
\cite{a:Seifert2000,a:ErdmannSchwarz2004b}.  Using the smallest value
for the loading rate, $m = 130 \pNs$, and the extracted bond
parameters $k_0 = 0.02 \Hz$ and $F_0 = 7.83 \pN$, we get a value of
$830$, which is much larger than the largest value $N_t^m = 10$ used
here. In principle, multiple bond experiments also offer the chance to
measure rebinding rates, in marked contrast to single bond
experiments. This could be done in different ways. For example, one
could control the time allowed for bond formation prior to loading
\cite{a:Zhu2000}. The mean number of formed bonds could be extracted
as described here and then converted into a value for the rebinding
rate. Alternatively, one could work at small loading rate. Then the
master equation from \eq{MasterEquation} had to be extended by
rebinding terms \cite{a:ErdmannSchwarz2004b}. A similar procedure as
described here then could be used to extract the rebinding rate.  For
biotin-streptavidin, the above estimate suggests that then loading
rates had to be smaller by at least one order of magnitude.  Therefore
adhesion bonds with faster off-kinetics than biotin-streptavidin might
be more appropriate for this purpose.

Finally, investigating multiple bond rupture will help to understand
properties of biological adhesion sites which usually consist of
clusters of adhesion molecules. For example, it has been found with
image correlation microscopy that in living cells during cell migration,
integrin adhesion receptors are preclustered with an average cluster
size of three to four, which is very similar to the average cluster size
studied here \cite{c:wise04}. Thus biomimetic studies like the one
presented here are essential to understand the way biological systems
make use of adhesion clusters.

This work was supported by the German Research Foundation (DFG)
through the Emmy Noether Program, the Center for Modeling and
Simulation in the Biosciences (BIOMS) at Heidelberg and the Human
Frontier Science Program (RG\#52/2003). Support from Corning
Corporation through the fellowship provided to one of us (S.P.) is
also gratefully acknowledged.

\end{document}